\documentclass[11pt]{article}
\usepackage{epsfig}
\usepackage{latexsym}
\begin{document}
%%%%%%%%%%%%%%%%%%%%%%%%%%%%%%%%%%%%%%%%%%%%%%%%%%%%%%%%%%%%%%%%%%%%%%%%%%%
\thispagestyle{empty}
\begin{flushright}
hep-lat/0107016
\end{flushright}
\begin{center}
\vspace*{6mm}
{\huge Properties of near-zero modes 
\vskip2mm
and chiral symmetry breaking}
\vskip13mm
{\bf Christof Gattringer${}^\dagger$, Meinulf G\"ockeler, P.E.L.~Rakow, \\
Stefan Schaefer and Andreas Sch\"afer}
\vskip5mm
Institut f\"ur Theoretische Physik \\
Universit\"at Regensburg \\
93040 Regensburg, Germany
\vskip28mm
\begin{abstract}
We study localization and chirality properties of eigenvectors
of the lattice Dirac operator. In particular we focus on the
dependence of our observables on the size of the corresponding
eigenvalue, which allows us to 
study the transition of a near-zero mode into a 
bulk mode. We analyze ensembles of quenched SU(3) 
configurations using a Dirac operator which 
is a systematic expansion in path length
of a solution of the Ginsparg-Wilson equation.
Our results support the interpretation of the excitations relevant for
chiral symmetry breaking as interacting instantons and 
anti-instantons.
\end{abstract}
\vskip3mm
{\sl To appear in Nuclear Physics B.}
\end{center}
\vskip6mm
\noindent
PACS: 11.15.Ha \\
Key words: Lattice QCD, instantons, chiral symmetry breaking
\vskip5mm \nopagebreak \begin{flushleft} \rule{2 in}{0.03cm}
\\ {\footnotesize \ 
${}^\dagger$ Supported by the Austrian Academy of Sciences (APART 654).}
\end{flushleft}
\newpage

\setcounter{page}{1}
\section{Introduction}
Chiral symmetry breaking is one of the most intriguing features 
of QCD and the existence of a non-vanishing chiral condensate is
a cornerstone in our understanding of hadron phenomenology. 
Analyzing and understanding the mechanisms which lead to the formation
of the chiral condensate thus can provide deep insights into the 
nature of relevant excitations of QCD.

In the last 25 years a phenomenological picture of chiral symmetry
breaking through the interaction of instantons and anti-instantons 
has been developed (see e.g.~\cite{SchSh98} for recent reviews). A single 
classical instanton or anti-instanton leads to a zero mode of the
Dirac operator. An infinitely separated instanton anti-instanton 
pair has two zero modes, but as soon as the fermion wave functions
start to overlap they mix, and the Dirac operator acquires
a pair of small complex conjugate eigenvalues. As more instantons and
anti-instantons are added they build up a non-vanishing 
density of eigenvalues near the origin.  
This non-vanishing density of eigenvalues in turn is related
to the chiral condensate via the Banks-Casher relation \cite{BaCa80}. 

This picture of chiral symmetry breaking through instantons provides a
wealth of signatures for the properties of the 
eigenvectors of the Dirac operator
with eigenvalues close to 0, the so-called {\sl near-zero modes}.  
In particular they should 
have pronounced localization and chirality properties.
These signatures can be directly tested in an ab-initio calculation 
using lattice gauge theory. The recent progress in 
understanding chiral symmetry on the lattice based on the Ginsparg-Wilson 
relation \cite{GiWi82} now
allows us to probe the QCD vacuum with a Dirac operator which respects 
chiral symmetry. Analyzing properties of eigenvectors of the Dirac operator
to learn about instantons is a more direct approach than analyzing the
gauge fields with cooling methods. The eigenvectors directly reflect 
the relevant excitations as seen by the Dirac operator without introducing
a filtering process by hand. Further evidence for the dominance of
instantons has been obtained by showing that fermionic $n$-point
functions can be saturated using spectral decomposition with 
only the lowest eigenmodes of the lattice Dirac operator 
\cite{saturate,degrandha,chuetal}.

Let us stress that while the instanton anti-instanton scenario of chiral 
symmetry breaking is highly popular, it is still far from being 
established. For an alternative based on monopole physics see
e.g.~\cite{zakharov} and references therein. Thus detailed numerical tests 
based on chirally improved lattice calculations are highly appropriate. 

In this article we present our results for locality and chirality 
properties of eigenvectors of the lattice Dirac operator. 
In particular we also focus on the dependence of our observables
on the size of the corresponding eigenvalue. This allows us to study 
how localization and chirality of a near-zero mode change as the
corresponding eigenvalue moves into the bulk of the spectrum. We analyze 
ensembles of quenched SU(3) configurations for different lattice sizes and
three values of the gauge coupling. We use a Dirac operator which is a 
systematic approximation of a solution of the Ginsparg-Wilson equation. It has 
very good chiral properties and is numerically cheaper than an exact solution
of the Ginsparg-Wilson equation such as e.g.~the overlap operator. Our results 
clearly support the interpretation of the excitations relevant for 
chiral symmetry breaking as interacting instantons and anti-instantons.

The article is organized as follows: In the next section we collect technical
preliminaries, in particular we discuss our gauge ensembles, the Dirac
operator we use and the details of the diagonalization. This is followed
by sections where we discuss our results for the spectral density 
(Section 3) of the Dirac operator and for the
localization properties of its eigenvectors (Section 4). In 
Section 5 we analyze the chirality properties of the near-zero modes. The
article closes with a summary of our results.

\section{Technical remarks}
\noindent
{\bf Gauge configurations:} We use ensembles of SU(3) gauge fields
in the quenched approximation
generated with the L\"uscher-Weisz action \cite{LuWeact,Aletal95}.
In addition to the standard plaquette term with coefficient 
$\beta_1$ this action also contains a rectangle term with coefficient
$\beta_2$ and a parallelogram term with coefficient $\beta_3$ (see 
\cite{Aletal95} for a detailed discussion of the action). The coefficient
$\beta_1$ is an independent parameter, while $\beta_2$ and $\beta_3$ can
be computed from $\beta_1$ using e.g.~tadpole improved perturbation theory 
\cite{Aletal95,LeMa93}. We use $\beta_1 = 8.10,8.30$ and
8.45. The corresponding values of $\beta_2$ and $\beta_3$ were determined 
in \cite{Gaetal01}. 
We performed runs on lattices of size $8^4$, $12^4$
and $16^4$ and the statistics for our ensembles is given in 
Table \ref{gaugestat}. We use a mix of Metropolis and over-relaxation steps
to update the gauge fields.
In order to set the scale we computed the Sommer parameter
\cite{sommer} $r_0$ for the $16^4$ ensembles. We give our results for
$r_0/a$ together with the lattice spacing $a$ (assuming $r_0 = 0.5$\,fm)
in Table \ref{gaugestat}.

\begin{table}[ht]
\begin{center}
\begin{tabular}{c|ccc}
 & $\beta_1 = 8.10$ & $\beta_1 = 8.30$ & $\beta_1 = 8.45$ \\
\hline
$r_0/a$ & 3.94(16) & 4.67(13) & 5.00(5) \\
$a$ & 0.127(5)~fm & 0.107(3)~fm & 0.100(1)~fm \\
\hline
$8^4$  & 800 &  800 & 800 \\
$12^4$ & 400 &  400 & 400 \\
$16^4$ & 200 &  200 & 200 \\
\end{tabular}
\end{center}
\caption{Sommer Parameter $r_0$, lattice spacing $a$ and 
statistics for our gauge field ensembles.
\label{gaugestat}}
\end{table}
\noindent
\\
{\bf Dirac operator: } Our Dirac operator $D$ is a systematic expansion of
a solution of the Ginsparg-Wilson equation \cite{GiWi82}, 
\begin{equation}
\gamma_5 \; D \; \; + \; \; D \; \gamma_5 \; \; = \; \; 
D \; \gamma_5 \; D \; .
\label{giwi}
\end{equation}
It has been understood during the last few years, that solutions of
(\ref{giwi}) can be used to implement chiral symmetry on the lattice. 
The two known\footnote{We remark that when one adds a fifth dimension
also the domain wall fermions provide a solution of the 
Ginsparg-Wilson equation.} exact solutions are perfect 
actions \cite{fixpd,Haetal00}
and overlap fermions \cite{overlap}. In a series of recent publications 
\cite{Gaetal01,GaHiLa01,GaHi00,Ga00} it has been demonstrated that for many
problems it is sufficient to work with an approximate solution of
(\ref{giwi}) which is considerably cheaper from a numerical point of view. 
The idea \cite{Ga00} is to expand the most general Dirac operator $D$ 
on the lattice into a sum of simple operators. Each of these simple operators
is built from a set of paths of the same length.
The members of these sets are 
related by symmetry transformations as e.g.~rotations, reflections etc. 
Along each path one builds a gauge 
transporter from the link variables and multiplies it with a Kronecker symbol 
with the endpoints of the path as arguments. The most general Dirac operator
is then obtained by multiplying this object with an element of the Clifford 
algebra and summing all the resulting simple lattice operators with some real 
coefficients \cite{Haetal00,Ga00}. 

Once the expansion of the most general Dirac operator is established it
can be inserted into the Ginsparg-Wilson equation. The left hand side 
can be evaluated easily by commuting the Clifford algebra elements in the
expansion with $\gamma_5$. The right hand side amounts to a commutation
with $\gamma_5$ and multiplications
of the individual terms in the expansion of $D$. It can be shown 
\cite{Ga00} that 
the product of any two terms of our expansion which appears on the
right hand side of (\ref{giwi})
gives a term which is already present in the full 
expansion of $D$, i.e.~shows up also on the left hand side
of (\ref{giwi}). The 
coefficients in front of the individual terms on both sides of
(\ref{giwi}) have to be equal. 
This results in a system of coupled quadratic
equations which can be solved numerically. In \cite{GaHiLa01} an approximate
solution 
with 19 terms was constructed and analyzed. We use the same solution here. 
A detailed description of the terms 
as well as the coefficients can be found in the
appendix of \cite{Gaetal01}. Our expansion of a solution of
the Ginsparg-Wilson equation is an expansion in terms of path 
length, since the size of the coefficients multiplying each 
term in the expansion decreases as the length of the paths increases
\cite{GaHiLa01}.

The symmetries that are implemented and are responsible for grouping the
paths into the above mentioned sets are lattice translations and 
lattice rotations as well as parity, charge conjugation and 
$\gamma_5$-hermiticity, where $\gamma_5$-hermiticity is defined as 
\begin{equation}
D \; \gamma_5 \; \; = \; \; \gamma_5 \; D^\dagger \; .
\label{g5herm}
\end{equation}
The property (\ref{g5herm}) can be seen 
\cite{Haetal00,Ga00} to correspond to the 
properties of $D$ which are used to 
prove CPT in the continuum. Eq.~(\ref{g5herm}) has an important implication
for the $\gamma_5$-matrix element of the 
eigenvectors of $D$. Let $\psi$ be such an eigenvector, 
i.e.~$D \psi = \lambda \psi$. Then a 
few lines of algebra show that \cite{itoh}
\begin{equation}
\psi^\dagger  \; \gamma_5 \; \psi \; \; = \; \; 0 \; \; , \; \; 
\mbox{ unless } 
\;  \lambda \; \mbox{ is real } \; .
\label{g5sandwich}
\end{equation}
This equation is the basis for the identification of topological 
modes. For the 
continuum Dirac operator and for a $D$ which is
an exact solution of the Ginsparg-Wilson 
equation one has exact zero modes.
Only zero modes have a non-vanishing matrix element with $\gamma_5$.
Here we are working with an approximate solution of the Ginsparg-Wilson
equation and we do not have exact zero modes. However, eigenvectors
for which the corresponding eigenvalue has a non-vanishing 
imaginary part cannot be candidates for zero modes, since their
matrix elements with $\gamma_5$ vanish exactly. In our approximation the
topological modes show up as eigenvectors with a small real eigenvalue.
For the real eigenvalues on the $16^4$ lattices we find mean values of
0.0124, 0.0122 and 0.0095 for $\beta_1 = 8.10$, $\beta_1 = 8.30$ and
$\beta_1 = 8.45$ respectively. The distributions (see \cite{Gaetalnew})
of the real eigenvalues are very narrow and we find 
that 95\% of the real eigenvalues lie between 0 and 0.05. 
The matrix elements $\psi^\dagger \gamma_5 \psi$ typically
have a value $\sim \pm 0.8$. If one added more 
terms in our approximation then the real modes would be closer to $0$ 
and the matrix element with $\gamma_5$ closer to $\pm 1$. 
\\
\\
{\bf Computation of the eigenvalues: }
For the computation of the eigenvalues and 
eigenvectors we used the implicitly
restarted Arnoldi method \cite{arnoldi}.
For lattice sizes $8^4$ and $12^4$ we computed 50 eigenvalues and the 
corresponding eigenvectors for each of our gauge field 
configurations and 30 eigenvalues and eigenvectors for the ensembles
on the $16^4$ lattice. The search criterion for the eigenvalues 
was their modulus, i.e.~we 
computed eigenvalues in concentric circles around the 
origin until 50, respectively 30 were found. The boundary conditions for 
the fermions were periodic in the space directions and anti-periodic 
in time. 

\section{Density of eigenvalues}

When describing chiral symmetry breaking based on instanton phenomenology, 
a major ingredient is the Banks-Casher formula
\cite{BaCa80} which 
relates the density $\rho(\lambda)$ of eigenvalues $\lambda$ at the origin
to the chiral condensate,
\begin{equation}
\langle \overline{\psi} \psi \rangle \; \; = \; \; - \pi \; \rho(0) \; 
V^{-1} .
\label{baca}
\end{equation}
We use a convention for the density $\rho(\lambda)$ 
such that $\rho(\lambda)\, \Delta \lambda$ gives the
number of eigenvalues in the interval $\Delta \lambda$.
The second ingredient for chiral symmetry breaking through instantons is
the above mentioned interaction of pairs 
of instantons and anti-instantons  
which leads to a pair of small complex conjugate eigenvalues.
In the chirally broken phase of QCD the abundance of interacting
instantons and anti-instantons 
builds up a finite density of eigenvalues near the origin thus
leading to a non-vanishing chiral condensate. This picture provides a 
wealth of signatures for the eigenvectors of the Dirac operator with small 
eigenvalues, the near-zero modes.

\begin{figure}[p]
\begin{center}
\epsfig{file=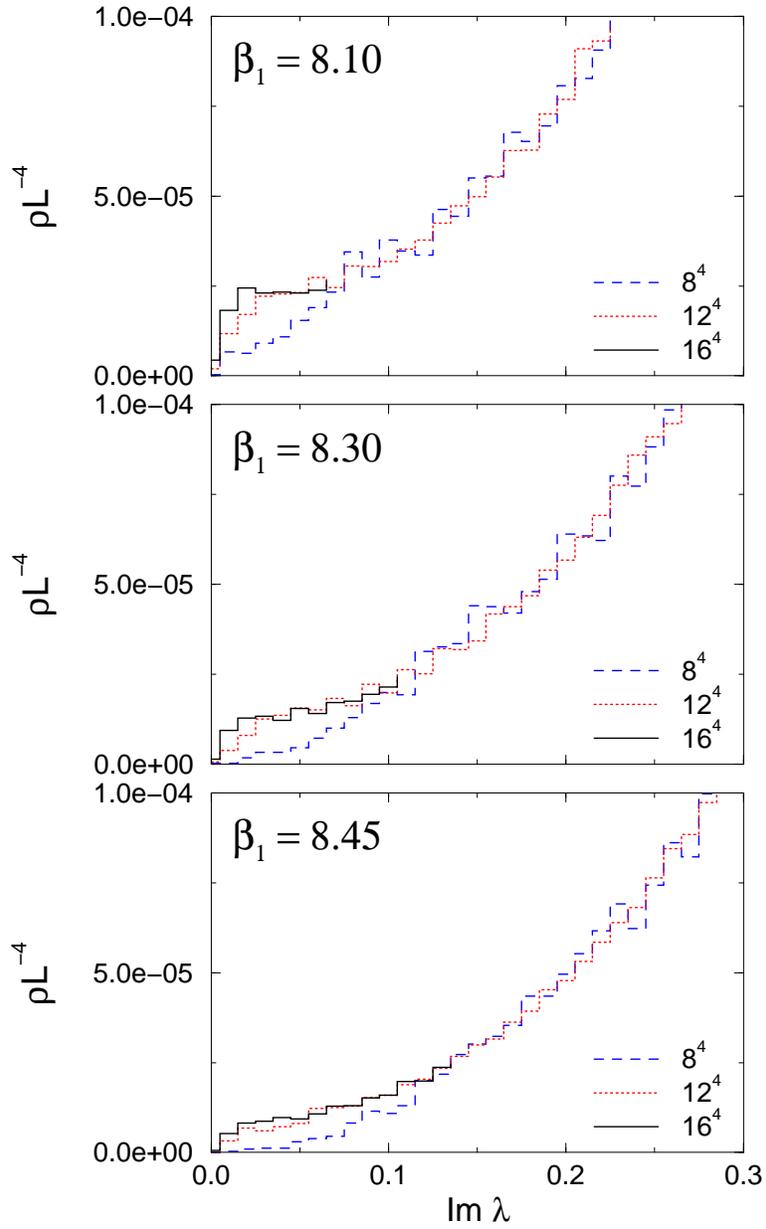,width=10cm,clip}
\caption{Density of eigenvalues as a function of $\mbox{Im}\,\lambda$.
The density is normalized by the inverse volume
of the corresponding lattices. Zero modes were left out for the
evaluation of the density.
\label{valdens}}
\end{center}
\end{figure}

Before we discuss our tests of some of
these characteristic features of the near-zero 
modes we briefly present our results for the density of eigenvalues. 
For the continuum Dirac operator the spectrum is
restricted to the imaginary axis and the definition of
the spectral density is straightforward. For a solution of 
the Ginsparg-Wilson equation (\ref{giwi}) 
the spectrum is located on a circle of radius 1 with center 1 in the 
complex plane. For our approximate solution $D$ of (\ref{giwi}) 
the spectrum fluctuates slightly around this circle (compare e.g.~the
spectra shown in \cite{Gaetal01}). The non-vanishing real parts of
the eigenvalues of an exact or approximate Ginsparg-Wilson fermion 
require a small modification in the definition of the eigenvalue density 
$\rho(\lambda)$. We bin the eigenvalues with respect to 
their imaginary part and count the number of eigenvalues
in each bin. Eigenvalues with vanishing imaginary part, i.e. the
zero modes, were left out.
After dividing the count in each bin by the total number
of eigenvalues we obtain the histograms for $\rho(\lambda)$ used in 
Fig.~\ref{valdens} below. Since we are only interested in the
density $\rho(\lambda)$ in a region of $\lambda$ 
where the real parts of the eigenvalues on
the Ginsparg-Wilson circle are small, binning with respect to the
imaginary parts of $\lambda$ gives a good approximation of
the continuum definition of the spectral density.

In Fig.~\ref{valdens} we show our results for the density of eigenvalues. 
We use the spectra for $\beta_1 = 8.10$ (top plot), $\beta_1 = 8.30$
(middle plot) and $\beta_1 = 8.45$ (bottom plot) and display
the results for all three volumes 
$8^4$ (dashed curve) $12^4$ (dotted curve) and $16^4$ (full curve). 
We remark that for the largest lattice ($16^4$) the 
smallest Matsubara frequency
has a value of $\pi/16 = 0.1963$, such that many of the eigenvalues are 
considerably below the smallest eigenvalue for the free case.
We plot the density $\rho(\lambda)$ normalized by the volume $L^4$
as a function of  
$\mbox{Im}\,\lambda$ 
where, since the curve is symmetric with respect to reflection 
at the origin, we display only positive $\mbox{Im}\,\lambda$. 
One finds that the data for the density essentially fall on 
universal curves. Only near the origin the density is more rounded  
for the smaller lattices in agreement with
universal random matrix theory predictions \cite{randommat}. For the
same reason also the curve for the largest system ($16^4$) still
shows a drop near the origin. Up to this finite-size effect, the density
remains non-zero down to $\lambda = 0$, thus building up the   
chiral condensate according to the Banks-Casher formula (\ref{baca}). We 
remark that at $\beta_1 = 8.45$ the smallest lattice ($8^4$) already 
is in the deconfined phase, i.e.~there the density develops a gap
(compare \cite{Gaetal01}). It is obvious that for $\beta_1 = 8.10$,
i.e.~the ensemble with the largest value of the gauge coupling
the density 
of eigenvalues and thus the condensate is largest while it becomes smaller
with increasing $\beta_1$.

We remark that in a recent study \cite{edwardsgap} 
of quenched QCD at finite temperature using the overlap operator,
a small nonvanishing density of eigenvalues near the origin was found 
even in the deconfined phase. In our recent paper \cite{Gaetal01} with the
ultralocal, chirally improved operator no
statistically significant signal for a nonvanishing density in 
the deconfined phase was seen.
This discrepancy might be related to an observation we made recently
\cite{Gaetal01c}. For
smooth instantons on the lattice the large extent of the
overlap operator considerably distorts the zero modes for instantons
with radius $\le 2.5$ in lattice units, i.e.~objects with a diameter 
of $\le 5.0$. We thus suspect
that using the overlap operator on lattices with time extent
4 (as in \cite{edwardsgap}) may lead to serious finite size effects.

\section{Localization properties}

Let us now come to the analysis of the characteristic features of the 
near-zero modes, i.e.~the eigenvectors of the Dirac operator with small
eigenvalues. In order to develop these characteristic features we first 
discuss the behavior of exact zero modes in the continuum. 
For an instanton configuration in the continuum, the Dirac operator has
an exact zero mode $\psi$ which is localized in space-time
around the center of the instanton 
and this localization has a radius proportional to that
of the underlying instanton. 
A gauge invariant density which displays this localization is
obtained by summing $|\psi(x,d,c)|^2$ over the Dirac and color 
indices $d$ and $c$ at each space-time point $x$ individually. 
A lattice discretization of this scalar density $p(x)$ is given by
\begin{equation}
p(x) = \sum_{c,d} \; \psi(x,c,d)^* \;  \psi(x,c,d) \; ,
\label{densdef}
\end{equation}
where now $\psi(x,c,d)$ are the entries of the eigenvectors of our 
lattice Dirac operator $D$. As expected from the analytic result in 
the continuum, this density shows a clear localization
at the position of an isolated smooth instanton put on the lattice by hand
\cite{Gaetal01c,clinst}. 
We remark that since the eigenvectors $\psi$ of $D$ are
normalized one has 
\begin{equation}
\sum_x \; p(x)  \; \; = \; \; 1 \; .
\label{denssum}
\end{equation}
A zero mode of the continuum
Dirac operator is also an eigenstate of $\gamma_5$. Hence it is
interesting to also analyze the gauge invariant pseudoscalar density $p_5(x)$ 
defined by
\begin{equation}
p_5(x) = \sum_{c,d,d^\prime} \; \psi(x,c,d)^* \; 
(\gamma_5)_{d,d^\prime} \; \psi(x,c,d^\prime) \; .
\label{dens5def}
\end{equation}
When evaluating the corresponding continuum quantity for a
zero mode, $p_5(x)$ differs from its scalar counterpart  
only by a possible sign, i.e.~the scalar density equals the 
pseudoscalar density for an instanton but the two densities have opposite
sign for an anti-instanton. 
If one puts a smooth instanton anti-instanton pair 
on the lattice by hand the density of the corresponding near-zero mode
shows a {\sl dipole behavior}. Near the instanton peak
of the pair $p_5(x)$ is positive while it is negative near
the center of the anti-instanton.
When analyzing the densities $p(x)$ and $p_5(x)$ for
thermalized configurations one still finds localized 
structures and also dipole structures which can be interpreted as 
signatures of isolated instantons or instanton anti-instanton
pairs (see e.g.~\cite{Gaetal01}
for plots of examples). Using cooling techniques to identify the
instantons independently it has been found that the localization 
of the eigenvectors is concentrated at the same region where the cooling
procedure finds an instanton (see e.g.~\cite{chuetal}). 

In order to go beyond analyzing the localization properties of
the eigenvectors by merely inspecting plots of the scalar 
density it is convenient to introduce the scalar  
inverse participation ratio $I$, which is widely used
in solid state physics,
\begin{equation}
I \; \; = \; \; V \; \sum_x \; p(x)^2 \; .
\label{idef}
\end{equation}
$V$ denotes the volume $L^4$. From its definition in 
Eq.~(\ref{densdef}) it follows that $p(x) \ge 0$ for all $x$. Taking into
account the normalization Eq.~(\ref{denssum}) one finds that a maximally
localized eigenvector which has support on only a single site $x^\prime$
must have $p(x) = \delta_{x,x^\prime}$. Inserting this into the definition
(\ref{idef}) for the inverse participation ratio one finds that a 
maximally localized eigenvector has $I = V$. Conversely, a maximally
spread eigenvector has $p(x) = 1/V$ for all $x$. In this case, the
inverse participation ratio gives a value of $I = 1$. Thus $I$ has a high
value for localized states while $I$ is small for delocalized ones. 
An alternative measure for the localization of the eigenvectors,
based on their self correlation has been studied in \cite{degrandha}.

We now use the inverse participation ratio to 
analyze the localization of the near-zero modes. According to the 
instanton picture the small eigenvalues come from interacting
instantons and anti-instantons. If an instanton and an anti-instanton
are infinitely far apart, i.e.~they do not interact, 
the Dirac operator still sees them as two independent objects and displays two
eigenvalues 0. As soon as the two objects interact this
degeneracy is lifted and the two eigenvalues split into a pair of
small complex conjugate eigenvalues. When the instanton and the anti-instanton 
approach further, the two eigenvalues move further up (down) in imaginary 
direction. At the same time the instanton and the anti-instanton start 
to annihilate. The localization is washed out
and the corresponding eigenvalues end up in the bulk of the
spectrum. Thus one expects that for eigenvalues very close to the 
origin the localization is still large since the two partners
remain relatively unperturbed. On the other hand eigenvalues further up (down)
the imaginary axis come from pairs where the partners are already
relatively close to each other and should be less localized.

\begin{figure}[t]
\begin{center}
\hspace*{-5mm}
\epsfig{file=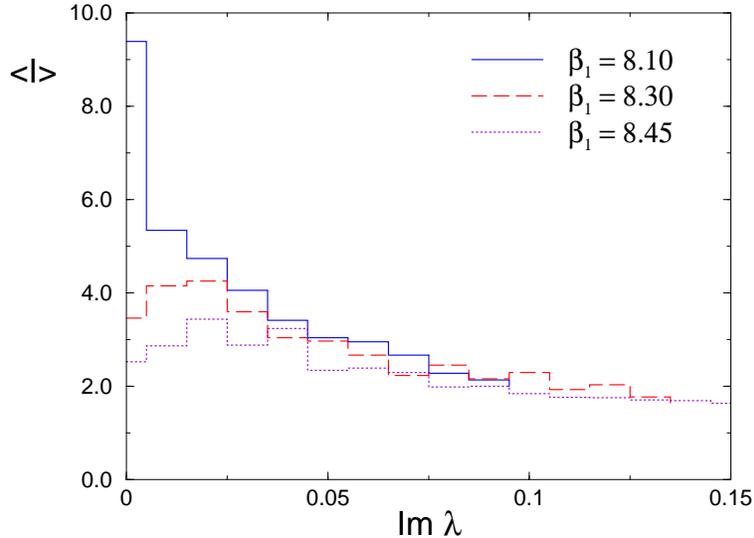,width=10cm,clip}
\caption{The average inverse participation ratio $\langle I \rangle$
of the near-zero modes as a function of $\mbox{Im}\,\lambda$
for lattice size $16^4$.
We remark that in the first bin real modes were left out.
\label{iprvslambda}}
\end{center}
\end{figure}

\begin{figure}[t]
\begin{center}
\hspace*{-5mm}
\epsfig{file=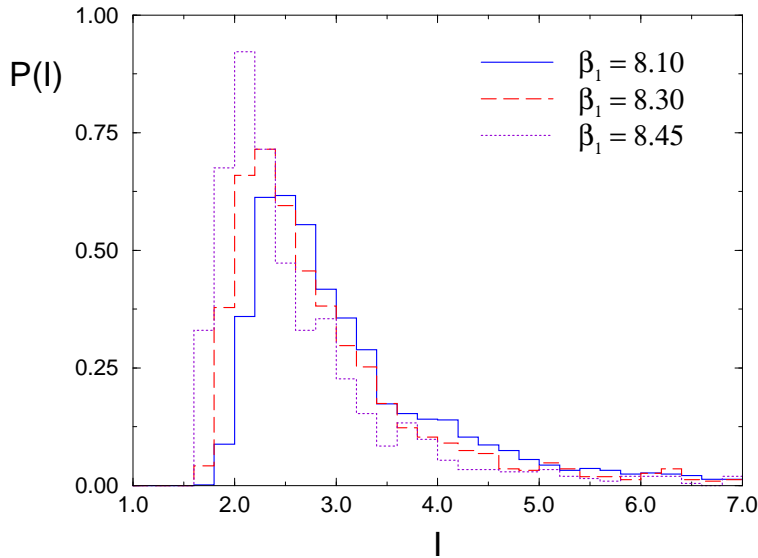,width=10cm,clip}
\caption{Distribution of the inverse participation ratio for the
near-zero modes with $0 < |\,\mbox{Im}\,\lambda\,| \leq 0.05$.
\label{iprdist}}
\end{center}
\end{figure}

In Fig.~\ref{iprvslambda} we plot our results for the average 
$\langle I \rangle$ of the inverse participation
ratio as a function of $\mbox{Im}\,\lambda$. The histograms were obtained by
binning $\mbox{Im}\,\lambda$
and computing the average of $I$ for the eigenvectors 
with eigenvalues in each bin. Topological modes, i.e.~modes with small
real eigenvalues were left out in the computation 
of the histograms\footnote{We remark 
that the topological modes have values of $I$
varying between 1.8 and up to 100, with the highest values of $I$ 
coming from defects. The distribution is peaked
near small values similar to the distribution for the near zero modes
shown in Fig.~\ref{iprdist}.}. Since the distribution is symmetric
with respect to the origin we display the curves only for 
$\mbox{Im}\,\lambda > 0$.
We show our results for the $16^4$ lattice at all three values of 
$\beta_1$. All three curves display their maximum at or near the origin,
as expected from the picture of interacting instantons and anti-instantons.
The same behavior was observed for the near-zero modes of the 
staggered Dirac operator in \cite{regensburg}.
Furthermore the eigenvectors for the $\beta_1 = 8.10$ ensembles have the
largest average of $I$ near $\lambda = 0$
while the eigenvectors for $\beta_1 = 8.45$ have 
the smallest values. This poses an interesting question: 
Is it simply a larger
number of localized states which leads to a larger condensate for 
$\beta_1 = 8.10$, or do in addition also the near zero modes themselves 
become more localized? This question can be analyzed by studying the
probability distribution of the inverse participation ratio.

In Fig.~\ref{iprdist} we show the probability distribution $P(I)$ 
of the inverse participation ratio for the near-zero modes 
(no zero modes were taken into account).
More precisely, we show the
probability distribution for all eigenvectors with eigenvalues obeying
$0 < |\mbox{Im}\;\lambda| \leq 0.05$, i.e.~for those modes for which
the curves in Fig.~\ref{iprvslambda} show a strong 
$\beta_1$ dependence. The distributions $P(I)$ are
normalized such that their integral over $I$ equals 1. It is obvious from 
Fig.~\ref{iprdist}, that for $\beta_1 = 8.10$ the distribution is shifted 
towards larger values of $I$, while it moves towards smaller values of $I$
as $\beta_1$ is increased. Thus not only the
average $\langle I \rangle$ of the inverse participation ratio is larger
for smaller $\beta_1$, but also the probability of finding a localized 
mode increases with smaller $\beta_1$. 

So far we have only concentrated on the localization of the
near-zero modes. According to the instanton picture of chiral symmetry
breaking also their local chiral behavior
should have an interesting signature. In particular $p_5(x)$ should 
have positive sign near instanton peaks and 
negative sign near anti-instanton
peaks. In the next section we will
discuss an observable introduced in \cite{locchir} which tests 
the local chirality for each lattice point individually. Before we come 
to our results for the local chirality let us first look at an observable 
which is a global measure for the chirality of an eigenvector. 

Similarly to the scalar inverse participation ratio $I$ of 
Eq.~(\ref{idef}) we can also define a measure $I_5$ 
for the localization of the
pseudoscalar density $p_5(x)$ (see Eq.~(\ref{densdef})),  
\begin{equation}
I_5 \; \; = \; \; V \; \sum_x \; p_5(x)^2 \; .
\label{ipr5def}
\end{equation} 
Let us discuss the behavior of $I_5$ for a smooth instanton 
and a smooth instanton anti-instanton pair. 
For a zero mode, corresponding to a single
smooth instanton or anti-instanton, one has 
$I = I_5$ since for all $x$ the $p(x)$ differ from the $p_5(x)$ 
only by a global sign. 
For an instanton anti-instanton pair 
the pseudoscalar density $p_5(x)$ has to undergo a
change of sign, since for $x$ near the instanton peak $p_5(x)$ is positive
while it is negative near the center of the anti-instanton.
This implies that $p_5(x)$ has to go through 0 for some $x$. Since
(compare the definitions (\ref{densdef}),(\ref{dens5def}))
for each $x$ we have $|\,p_5(x)| \leq p(x)$, it follows from the definitions
of $I$ and $I_5$ that for an instanton anti-instanton pair 
one has $I_5 < I$. As long as the
two constituents are relatively remote from each other and much of their 
identity as an independent instanton, respectively anti-instanton 
remains intact
one still finds $I_5 \sim I$. However, 
for pairs where the partners are close 
to each other, one expects values of $I_5$ considerably smaller than $I$.
Thus one expects to find a relatively large ratio of $I_5/I$ for modes with
eigenvalues very close to the origin and a drop of this ratio
as $|\,\mbox{Im}\,\lambda|$ increases.

\begin{figure}[t]
\begin{center}
\vspace*{3mm}
\hspace*{-10mm}
\epsfig{file=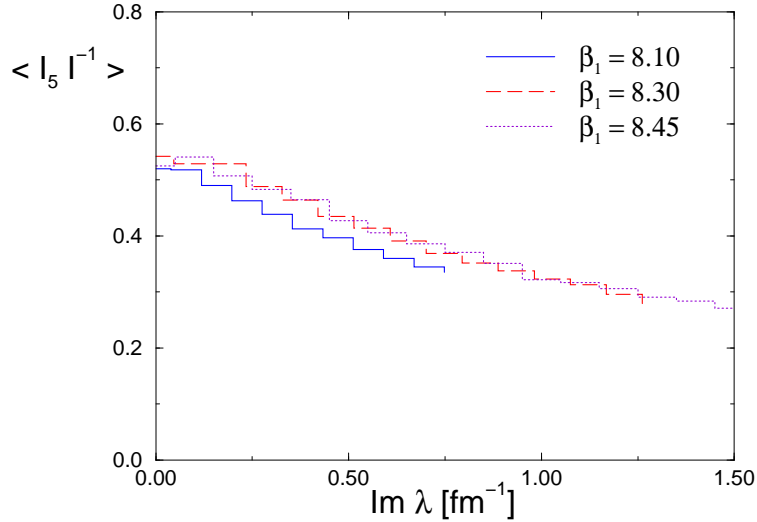,width=10cm,clip}
\caption{The ratio $I_5/I$ of pseudoscalar and scalar 
inverse participation ratio as a function of
$\mbox{Im}\,\lambda$ (in physical units). 
The data were computed on the $16^4$ lattices.
\label{i5iratvslambda}}
\end{center}
\end{figure}

In Fig.~\ref{i5iratvslambda} we show our results for the average
$\langle I_5/I \rangle$ as a function of $\mbox{Im}\,\lambda$. Again 
we do not take into account zero modes.
One clearly sees the 
drop of $\langle I_5/I \rangle$ as Im $\lambda$ increases. This is in 
agreement with the above discussed picture of instanton anti-instanton
pairs where the
two partners lose their chiral identity as they approach each other. 
Note that the curves for the three different values of $\beta_1$
are very similar.

\section{Local chirality}

To further study the nature of the near-zero modes we analyze 
the local chirality observable proposed by Horv\'ath et al.~in 
\cite{locchir}. In the original article \cite{locchir} the 
computation was done with Wilson fermions and the result was 
interpreted as evidence against the instanton picture. In the meantime
the local chirality variable was reanalyzed in several articles 
\cite{locchir2} using different Dirac operators
or a modification of the observable. These latter 
collaborations all interpret their results in favor of the instanton picture. 
Here we add further supporting evidence by analyzing for the first time 
how the local chirality of a near-zero mode changes as  
$|\mbox{Im}\,\lambda|$ of the corresponding eigenvalue increases. 
This allows us to directly see how an instanton and an anti-instanton 
lose their chirality and the corresponding near-zero mode turns into a bulk
mode.

Before
we present our numerical results we briefly repeat the definition of
the local chirality variable of \cite{locchir}.
In analogy to the densities $p(x)$ and $p_5(x)$ 
of Eqs.~(\ref{densdef}),(\ref{dens5def}) we can
define local densities $p_+(x)$ and $p_-(x)$
with positive and negative chirality, 
\begin{equation}
p_\pm(x) = \sum_{c,d,d^\prime} \; \psi(x,c,d)^* \; 
(P_\pm)_{d,d^\prime} \; \psi(x,c,d^\prime) \; ,
\end{equation}
where $P_\pm$ denote the projectors onto positive and negative 
chirality $P_\pm = (1 \pm \gamma_5)/2$. It is now interesting 
to analyze locally for each lattice point $x$ the ratio
$p_+(x)/p_-(x)$. For a classical instanton in the continuum, 
the corresponding
zero mode $\psi(x)$ has positive chirality and the density $p_+(x)$
is positive for all $x$ while $p_-(x)$ vanishes everywhere. Thus
the ratio of the two always gives $\infty$. For an anti-instanton the 
roles of $p_+$ and $p_-$ are exchanged and the ratio is always 0. 
When now analyzing the eigenvectors for an interacting instanton 
anti-instanton pair this property should still hold approximately
near the peaks. The ratio $p_+(x)/p_-(x)$ is expected to be
large for all $x$ near the instanton peak
of the pair and small for all $x$ near the anti-instanton peak. 
In a final step Horv\'ath et al.~map the two extreme values $\infty$ and 
$0$ of the ratio $p_+(x)/p_-(x)$ onto the two values $+1, -1$ 
using the arctangent. One ends up with the local chirality variable
$X(x)$ defined as 

\begin{equation}
X(x) \; = \; \frac{4}{\pi} \; 
\mbox{arctan} \, \left( \sqrt{\frac{ p_+(x) }{ p_-(x) }} \right) \; - 
\; 1 \; .
\end{equation}
If the eigenvectors 
for the near-zero eigenvalues correspond to instanton anti-instanton
pairs then the values of $X(x)$ should cluster near $+1$ and $-1$ 
when one selects lattice points $x$ near the peaks of the
scalar density $p(x)$. One can choose different values for the 
percentage of points $x$ used for the average
and here we will present results
for cuts of 1\%, 6.25\% and 12.5\%. This means that we 
average over those 1\% (6.25\%, 12.5\%) of all lattice points 
which support the highest peaks of $p(x)$. 
The smallest cut-off of 1\% will give 
the best signature since only the highest peaks which are not so
much affected by quantum fluctuations contribute. On the other
hand for such a small cut-off the results are not expected to
be very conclusive, since typically 
instanton models have a packing fraction of
instantons considerably larger than 1\% (see \cite{SchSh98}).
As already announced, we study the dependence of the chiral density 
of the eigenvectors on the imaginary part $\mbox{Im}\,\lambda$ 
of the corresponding 
eigenvalues. To do so we bin Im $\lambda$ and sum the
contributions $X(x)$ for each bin individually. In order 
to compare the different bins we normalize $X(x)$ such
that the area under the histogram is 1 for each of the $\lambda$-bins.   

\begin{figure}[p]
\begin{center}
\epsfig{file=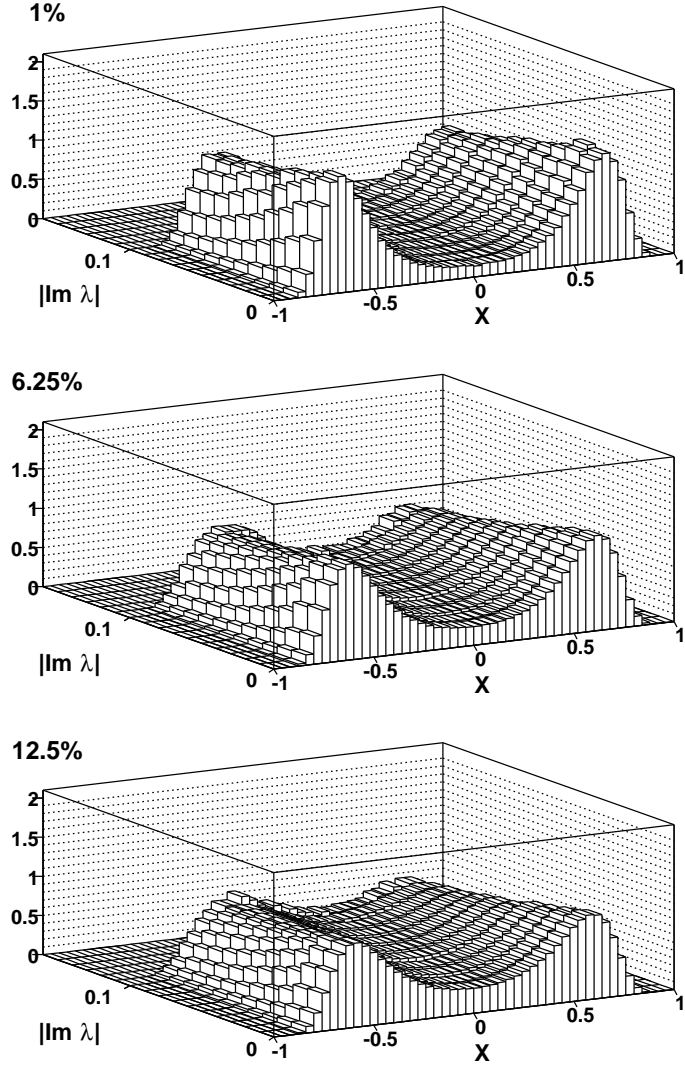,width=10.5cm,clip}
\caption{Local chirality for the near-zero modes in the $\beta_1 = 8.10$
ensemble on the $16^4$ lattices. We binned the data with respect to 
the imaginary part $| \mbox{Im}\,\lambda |$ of the eigenvalues
and normalized the histograms 
to 1 for each $\lambda$-bin individually. We use cuts of 1\% (top plot),
6.25\% (middle plot) and 12.5\% (bottom plot) 
for the percentage of supporting lattice points.
\label{locchir810}}
\end{center}
\end{figure}

\begin{figure}[p]
\begin{center}
\epsfig{file=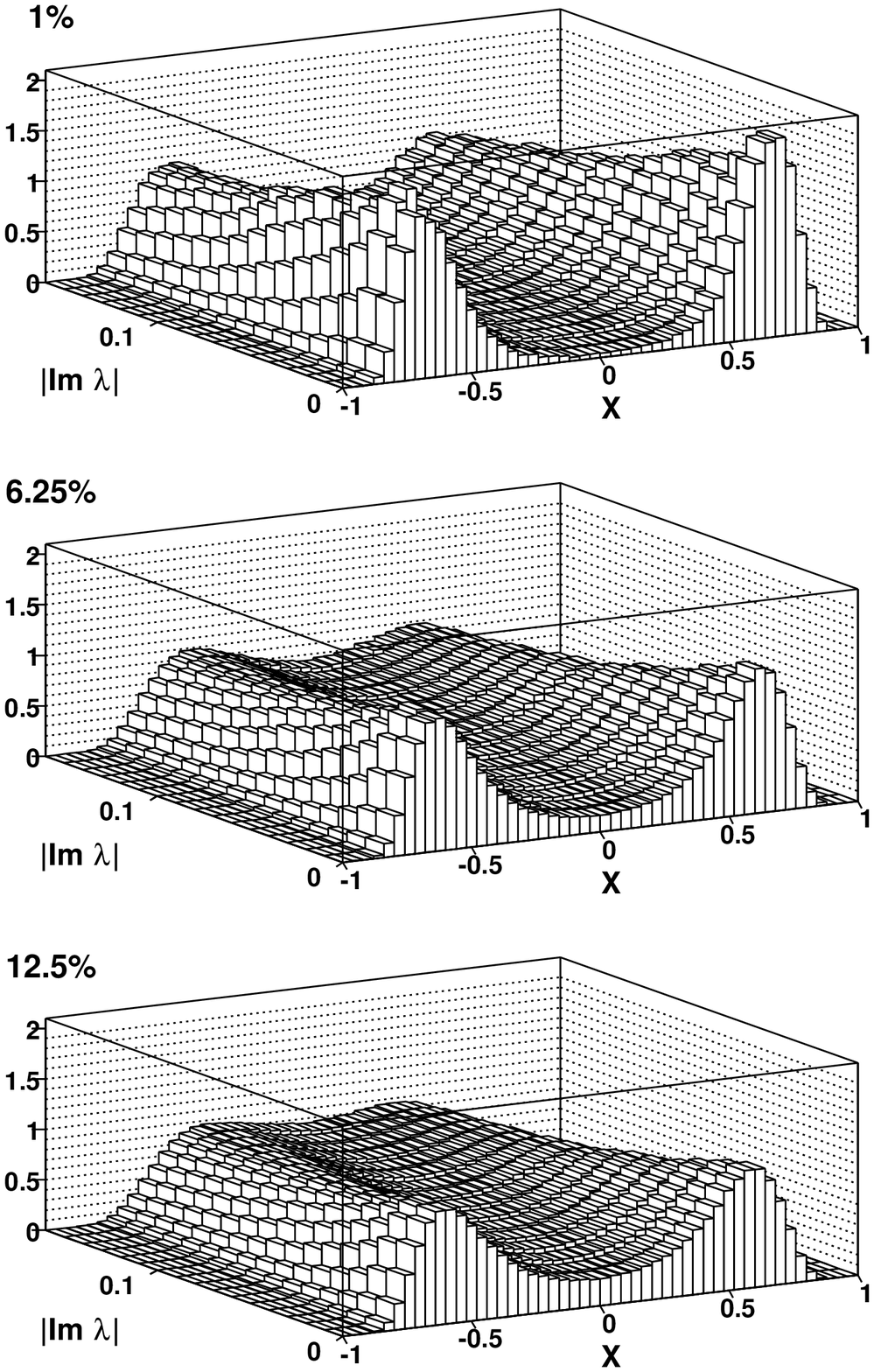,width=10.5cm,clip}
\caption{Local chirality for the near-zero modes as in Fig.~\ref{locchir810}
but now for $\beta_1 = 8.45$.
\label{locchir845}}
\end{center}
\end{figure}

In Fig.~\ref{locchir810} we show our results for the
$\beta_1 = 8.10$ ensemble on the $16^4$ lattice and Fig.~\ref{locchir845}
displays the local chirality for $\beta_1 = 8.45$. The top plots give
the result for the 1\% cut on the number of supporting lattice points, while
the middle and bottom plots are for 6.25\% and 12.5\% respectively. 
We remark that due to the higher density of eigenvalues $\lambda$ 
at $\beta_1 = 8.10$ our data which were computed from the eigenvectors
corresponding to the 30 lowest eigenvalues extend only to 
$| \mbox{Im}\,\lambda | = 0.1$ while we reach 
$| \mbox{Im}\,\lambda | = 0.15$ at $\beta_1 = 8.45$.
It is obvious, that for all plots the modes with eigenvalues closest 
to 0 have the best signal for local chirality, i.e.~the curves in
the first bin always show two well pronounced maxima in the vicinity 
of $\pm 1$ and a clear valley separating the two peaks. As 
$| \mbox{Im}\,\lambda |$ increases, the signal for local chirality weakens,
i.e.~the valley is less pronounced and the height of the peaks lowers. This 
reflects the washing out of the local chirality as
instantons and anti-instantons approach, as discussed above. 
The same effect has been observed 
for the global observable $\langle I_5 / I \rangle$ as shown in 
Fig.~\ref{i5iratvslambda}. Thus, as a near-zero mode turns into a bulk mode 
it loses its local chirality.

When comparing the results
for the different cuts on the percentage of the supporting
lattice points one finds, as expected, that the strongest
signal for local chirality is obtained for
the 1\%-cut where only the highest peaks are taken into account. However, 
also for cuts of 6.25\% and 12.5\% one still finds a clear signal for local
chirality.
Finally when comparing the results for the two different values of $\beta_1$
we observe in the $\beta_1 = 8.45$ ensemble a better signal of local 
chirality for the modes with eigenvalues very close to 0. This can be 
understood by the stronger suppression of quantum fluctuations at
larger values of $\beta_1$.

\section{Summary}
In this article we have studied localization and chirality properties 
of eigenvectors of the lattice Dirac operator with small eigenvalues. 
According to the picture of chiral symmetry breaking through instantons 
these near-zero modes should display characteristic features. In particular,
since they are supposed to trace interacting instantons and anti-instantons,
they are expected to be relatively localized and their pseudoscalar density
should display dipole behavior. We analyze these modes using the
scalar and pseudoscalar participation ratio and the local chirality
variable of Horv\'ath et al. In particular we focus on the
dependence of these observables on the imaginary parts $\mbox{Im}\,\lambda$
of the corresponding eigenvalues. One expects that both the locality 
and the local chirality of the near-zero modes decrease with increasing
$|\,\mbox{Im}\,\lambda\,|$. Our main findings are as follows:
\begin{itemize}

\item When plotting the average inverse participation 
ratio $\langle I \rangle$ as
a function of $\mbox{Im}\,\lambda$, we find the 
largest values of $\langle I \rangle$ near the origin and $\langle I \rangle$
decreases as $|\,\mbox{Im}\,\lambda\,|$ increases. 

\item A comparison of the probability distribution $P(I)$ of the
inverse participation ratio for the near-zero modes 
shows that for the strongest gauge coupling 
(i.e.~$\beta_1 = 8.10$), where the chiral condensate is biggest, one
has the largest probability of finding localized states.

\item A plot of the average ratio $\langle I_5/I \rangle$ as a function 
of $\mbox{Im}\,\lambda$ shows that the most chiral states are 
located near the origin and as $|\,\mbox{Im}\,\lambda\,|$ is
increased this chirality becomes washed out. We observe that the
ratio $\langle I_5/I \rangle$ falls on a universal curve for the
different values of $\beta_1$, i.e.~is relatively stable under
quantum fluctuations.

\item The observable of Horv\'ath et al.~gives a clear signal for 
local chirality for cuts of 1\%, 6.25\% and 12.5\% on the number of 
supporting lattice points. When $|\,\mbox{Im}\,\lambda\,|$
is increased we observe a decay of the local chirality for the
corresponding modes.

\end{itemize}

\noindent
Our results support the picture of the emergence of the chiral 
condensate through the interaction of instantons and anti-instantons.
\\
\\
{\bf Acknowledgements: } We would like to thank 
Peter Hasenfratz, Holger Hehl,
Ivan Hip, Christian B.~Lang, Ferenc Niedermayer,
Edward Shuryak, Wolfgang S\"oldner, Christian Weiss 
and Vladimir I.~Zakharov for
interesting discussions. This project was supported by 
the Austrian Academy of Sciences, the DFG and the BMBF. We thank 
the Leibniz Rechenzentrum in Munich for computer time on the Hitachi 
SR8000 and their operating team for training and support.

%\newpage

\end{document}